\documentclass[preprint,12pt]{elsarticle}
\usepackage{amsmath}
\usepackage{amssymb}
\usepackage{float}
\usepackage[T1]{fontenc}
\usepackage[utf8]{inputenc}
\usepackage{mathrsfs}
\usepackage[cmintegrals]{newtxmath}
\usepackage{multirow}
\usepackage{soul}
\soulregister\cite7
\soulregister\footnote7
\soulregister\eqref7
\soulregister\ref7
\usepackage{siunitx}
\usepackage[color=green!40]{todonotes}
\usepackage{threeparttable}
\usepackage{hyperref}
\journal{Computer Networks}

\graphicspath{{./}{figures/}}

\begin{document}

\begin{frontmatter}

  \title{On the Practical Implementation of Propagation Delay and Clock Skew
    Compensated High-Precision Time Synchronization Schemes with
    Resource-Constrained Sensor Nodes in Multi-Hop Wireless Sensor Networks}
	
  \author[xjtlu]{Xintao Huan}
  \author[xjtlu]{Kyeong Soo Kim\corref{cor}}
  \cortext[cor]{Corresponding author}
  \ead{Kyeongsoo.Kim@xjtlu.edu.cn}
  \address[xjtlu]{Department of Electrical and Electronic Engineering, Xi'an
    Jiaotong-Liverpool University, Suzhou 215123, Jiangsu Province, P. R.
    China}

  \begin{abstract}
    In wireless sensor networks (WSNs), implementing a high-precision time
    synchronization scheme on resource-constrained sensor nodes is a major
    challenge. Our investigation of the practical implementation on a real
    testbed of the state-of-the-art WSN time synchronization scheme based on the
    asynchronous source clock frequency recovery and the reverse two-way message
    exchange, which can compensate for both propagation delay and clock skew for
    higher precision, reveals that its performance on battery-powered,
    low-complexity sensor nodes is not up to that predicted from simulation
    experiments due to the limited precision floating-point arithmetic of sensor
    nodes. Noting the lower computational capability of typical sensor nodes and
    its impact on time synchronization, we propose an asymmetric high-precision
    time synchronization scheme that can provide high-precision time
    synchronization even with resource-constrained sensor nodes in multi-hop
    WSNs. In the proposed scheme, all synchronization-related computations are
    done at the head node equipped with abundant computing and power resources,
    while the sensor nodes are responsible for timestamping only. Experimental
    results with a testbed based on TelosB motes running TinyOS demonstrate that
    the proposed time synchronization scheme can avoid time synchronization
    errors resulting from the single-precision floating-point arithmetic of the
    resource-constrained sensor nodes and achieve microsecond-level time
    synchronization accuracy in multi-hop WSNs.
  \end{abstract}


\end{frontmatter}

\section{Introduction}
\label{sec-1}
High-precision time synchronization is essential to the collaborative
applications for wireless sensor networks (WSNs), including time-based channel
sharing and media access control (MAC) protocols \cite{huang13:_mac} and
coordinated duty cycling mechanisms \cite{carrano14:_survey}. Considering the
increasing number of WSN deployments for a variety of applications, most of
which are based on multi-hop topologies with resource-constrained sensor nodes,
achieving high-precision time synchronization in multi-hop networks while
lowering the computational requirements at sensor nodes is crucial in designing
WSN time synchronization schemes.

\begin{figure}[!tb]
  \centering \includegraphics[width=\linewidth]{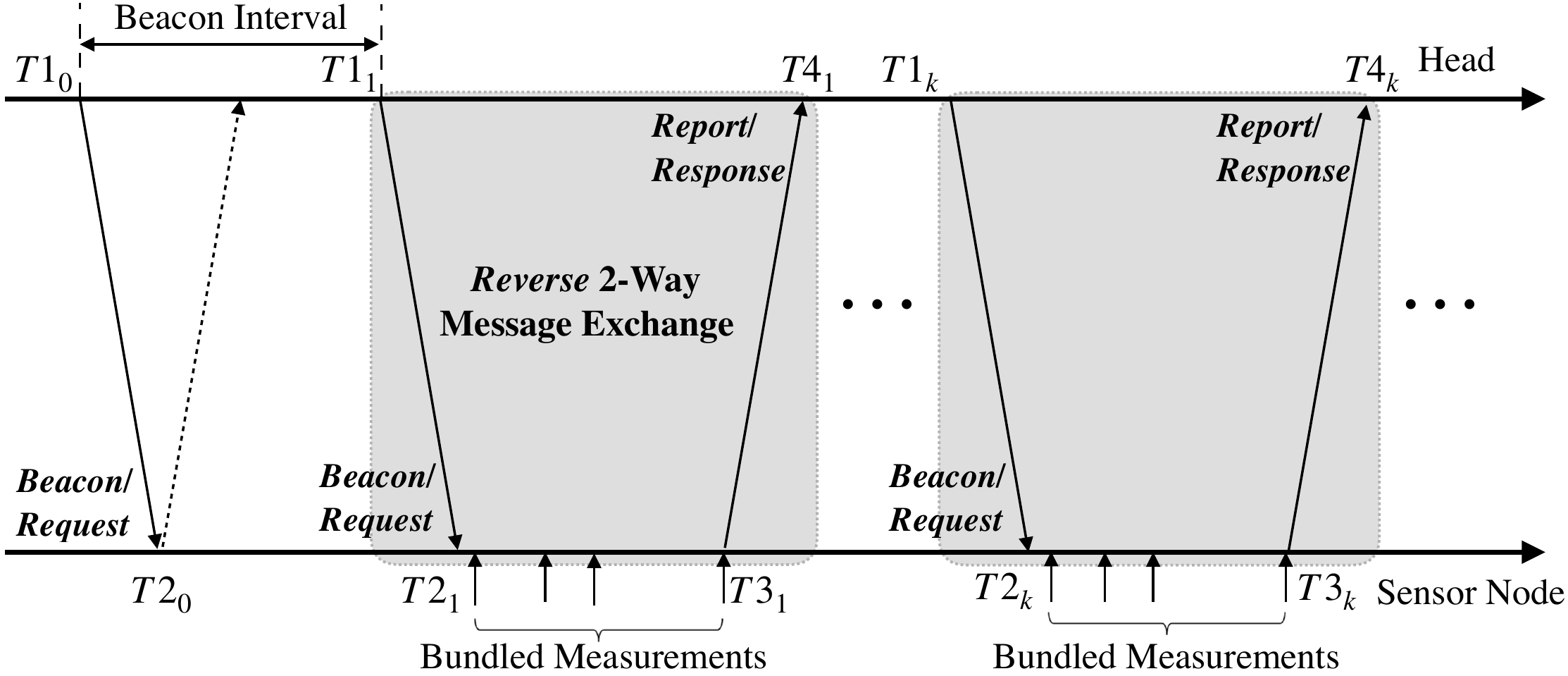}
  \caption{Reverse two-way message exchange with optional bundling of
    measurements introduced in \cite{Kim:17-1}.}
  \label{fig:reverse_twoway}
\end{figure}

In \cite{Kim:17-1}, we have proposed a novel energy-efficient time
synchronization scheme based on the asynchronous source clock frequency recovery
(SCFR) \cite{Kim:13-1} and the reverse two-way message exchange as illustrated
in Fig.~\ref{fig:reverse_twoway}, which we call \textit{EE-ASCFR} in short from
now on.
Unlike the conventional WSN time synchronization schemes---e.g., timing-sync
protocol for sensor networks (TPSN)
\cite{ganeriwal03:_timin_protoc_sensor_networ} and flooding time synchronization
protocol (FTSP) \cite{maroti04}---where sensor nodes are responsible for most of
the clock estimation procedures, EE-ASCFR suits the resource-constrained sensor
nodes, because sensor nodes are relieved from the task of clock offset
estimation that is moved to and done in a centralized manner at the head
node\footnote{A head node is also called a sink node in the literature.}; in
EE-ASCFR, the logical clocks of sensor nodes are synchronized to the reference
clock of the head node in frequency through the asynchronous SCFR, but they
could run possibly with different and independent offsets. Such redistribution
of the synchronization tasks reduces not only the message transmissions in the
two-way message exchange but also the computational complexity of sensor
nodes. Extensive simulation experiments demonstrate that EE-ASCFR can provide
sub-microsecond-level accuracy. Note that the actual performance of EE-ASCFR on
a real testbed was not evaluated at all in \cite{Kim:17-1}, which is the
starting point of our investigation reported in this paper.

To evaluate the actual performance of the high-precision time synchronization
schemes in practice, we implemented EE-ASCFR on TelosB \cite{telosb} motes
running TinyOS \cite{tinyos-doc} and investigated its time synchronization
performance on a real testbed. During the investigation, we found that the
limited computing capability of the sensor nodes could result in cumulative
synchronization errors in EE-ASCFR. This is because the estimation of the
frequency ratio and the maintenance of the logical clock require floating-point
divisions and the limited floating-point precision of the resource-constrained
sensor nodes---i.e., 32-bit single-precision on TinyOS---could result in
cumulative synchronization errors.

Note that, the 32-bit single-precision floating-point representation is the
floating-point standard not only of most resource-constrained WSN platforms such
as MicaZ \cite{MicaZ}, Iris \cite{Iris} and TelosB \cite{telosb} but also of
most Arduino platforms \cite{Arduino}. The latter platforms are extremely
prevalent in Internet of things (IoT) prototyping, and their computing and power
resources are also quite limited.  Moreover, the high-precision time
synchronization is also critical to collaborative IoT
applications. Consequently, minimizing the computing requirements of
high-precision time synchronization schemes for the resource-constrained
platforms is a timely and critical research topic for the success of WSNs and
IoT in the future.

Based on the results of the investigation of EE-ASCFR time synchronization
performance on a real WSN testbed, we propose an asymmetric high-precision time
synchronization (AHTS) scheme, which can still provide microsecond-level
accuracy even with resource-constrained sensor nodes by relieving them of all
time synchronization tasks but timestamping. We also present its multi-hop
extension to make it scalable in actual deployments.

The rest of this paper is organized as follows: The impact of the limited
precision floating-point arithmetic on time synchronization at
resource-constrained sensor nodes is investigated in detail in
Section~\ref{sec-2}. The proposed AHTS and its extension to multi-hop topologies
are described in Section~\ref{sec-3}. The results of experiments with a real
testbed for a comparative analysis of the performance of the proposed AHTS and
EE-ASCFR in both single-hop and multi-hop topologies are presented in
Section~\ref{sec-4}. Section~\ref{sec-5} concludes our work with directions for
future work.

\section{Impact of Limited Precision Floating-Point Arithmetic on Time
  Synchronization}
\label{sec-2}
We begin our investigation of the impact of limited precision floating-point
arithmetic on time synchronization at resource-constrained sensor nodes with
EE-ASCFR, the state-of-the-art time synchronization scheme proposed in
\cite{Kim:17-1}, which compensates for both propagation delay and clock skew to
provide sub-microsecond-level synchronization accuracy. In \cite{Kim:17-1}, the
performance of EE-ASCFR is evaluated based on mathematical analyses and
simulation experiments but not with a real testbed. To evaluate its actual
performance on the resource-constrained sensor nodes, therefore, we implemented
and evaluated it on a WSN testbed based on TelosB motes running TinyOS.

As discussed in Section~\ref{sec-1}, EE-ASCFR focuses on an asymmetric WSN with
a head node with higher computing and power resources and multiple
battery-powered, low-complexity sensor nodes. The asymmetric scenario of
EE-ASCFR represents the most common WSN applications such as environment
monitoring. As the hardware clocks of the sensor nodes are not ideal, they can
possibly have different clock frequencies and offsets with respect to the
reference clock.

In EE-ASCFR, the first-order affine clock model is used to model the hardware
clock $T_{i}$ of a sensor node $i$ with respect to the reference clock $t$ of
the head node \cite{Kim:17-1}: For $i{\in}\left[0,1,\ldots,N{-}1\right]$,
\begin{equation}
  \label{eq:hardware_clock_model}
  T_{i}(t) = \left(1+\epsilon_{i}\right)t + \theta_{i} ,
\end{equation}
where $N$ is the number of sensor nodes and $\epsilon_{i}{\in}\mathbb{R}$ and
$\theta_{i}{\in}\mathbb{R}$ denote the clock skew\footnote{A clock skew is
  defined as a normalized clock frequency difference between two clocks, and its
  typical value for clocks based on quartz crystal oscillators is of the order
  of tens of ppm (i.e., $\epsilon_{i}{\ll}1$) \cite{Kim:17-1}. Note that
  $(1+\epsilon_{i}){\in}\mathbb{R}_{+}$ in \eqref{eq:hardware_clock_model} is a
  clock frequency ratio.} and the clock offset between the reference clock and
the hardware clock of a sensor node $i$, respectively. Based on the hardware
clock $T_{i}$, the logical clock $\mathscr{T}_{i}$ used for timestamping at the
sensor node can be described as follows: For $t_{k}{<}t{\leq}t_{k+1}$
($k{=}0,1,\ldots$),
\begin{equation}
  \label{eq:logical_clock_model}
  \mathscr{T}_{i}\Big(T_{i}(t)\Big) = \mathscr{T}_{i}\Big(T_{i}(t_{k})\Big)
  + \dfrac{T_{i}(t)-T_{i}(t_{k})}{1 + \hat{\epsilon}_{i,k}} - \hat{\theta}_{i,k} ,
\end{equation}
where $t_{k}$ is the reference time for the $k$th synchronization,
$\hat{\epsilon}_{i,k}$ and $\hat{\theta}_{i,k}$ are the estimated clock skew and
offset from the $k$th synchronization. Note that $\hat{\theta}_{i,k}$ is set to
0 in \eqref{eq:logical_clock_model}, which is compensated at the head node as
described in \cite{Kim:17-1}; the sensor node only synchronizes the frequency of
the logical clock to that of the reference clock using asynchronous SCFR scheme.
According to the reverse two-way message exchange shown in
Fig.~\ref{fig:reverse_twoway}, the clock frequency ratio---i.e.,
$1{+}\epsilon_{k}$ in \eqref{eq:hardware_clock_model}---is estimated as
$(T2_{k}{-}T2_{0}){/}(T1_{k}{-}T1_{0})$, where the timestamps of $Ti_{j}$
($i{=}1,2$ and $j{\geq}0$) are recorded during the $j$th synchronization.

\begin{figure}[!tb]
  \newlength{\myvspace} \setlength{\myvspace}{0.3cm}
  \begin{center}
    \includegraphics[width=.85\linewidth]{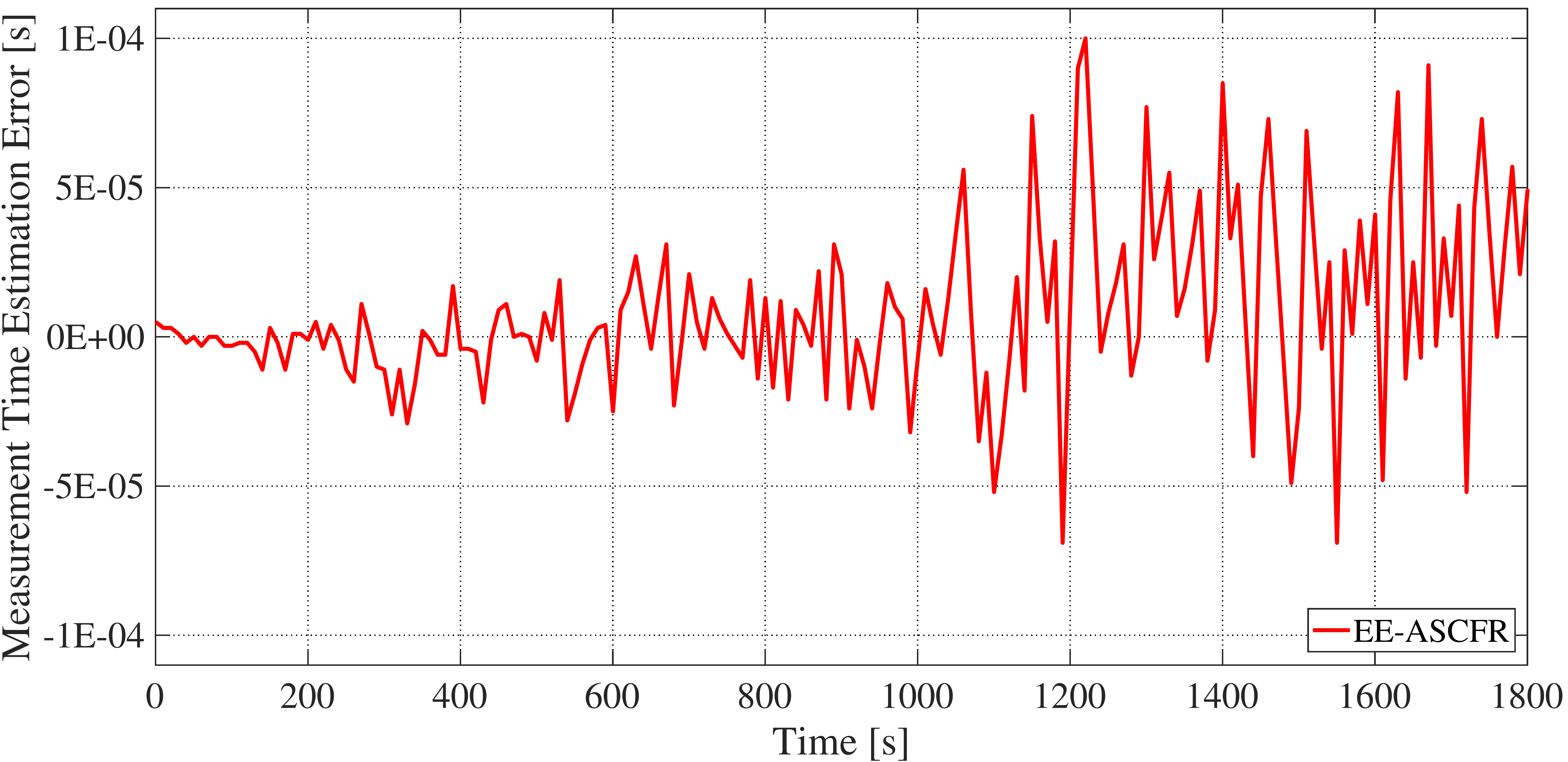}
  \end{center}
  \caption{Measurement time estimation errors of EE-ASCFR with SI of
    \SI{1}{\s}.}
  \label{fig:MTEoverTime}
\end{figure}

Fig.~\ref{fig:MTEoverTime} shows the measurement time estimation errors from the
experiment with the testbed for a period of \SI{1800}{\s}, where the
synchronization interval (SI) is set to \SI{1}{\s} and 5 measurements are
generated and bundled together in a ``Report/Response'' message to the head node
in each SI.
As shown in Fig.~\ref{fig:MTEoverTime}, the absolute value of measurement time
estimation error of EE-ASCFR gradually increases from around \SI{2}{\us} to
\SI{100}{\us} over the period of \SI{1200}{\s}, which indicates that, as will be
discussed shortly, the limited precision in floating-point arithmetic of the
resource-constrained sensor nodes (i.e., 32-bit single-precision in this case)
has negative impacts on the time synchronization performance.

Note that, when proposing the ratio-based time synchronization protocol (RSP)
\cite{rsp}, i.e., a variation of FTSP based on a simpler ratio-based clock
estimation method, the authors discuss the impact of computational errors
resulting from the limited precision floating-point arithmetic on time
synchronization in a qualitative way. Specifically, they claim that a smaller
synchronization time interval could lead to larger computational errors, while a
larger synchronization time interval, too, may negatively affect time
synchronization due to the clock drift over a long period. This means that we
need to address the impact of the computational errors due to the
limited-precision floating-point arithmetic in designing the high-precision time
synchronization schemes for the resource-constrained sensor nodes.

To systematically investigate the cause of the increase of the measurement time
estimation errors over time in EE-ASCFR, we revisit the arithmetic computations
involved with the logical clock updates in \eqref{eq:logical_clock_model} at the
sensor node. During the simulation experiments reported in \cite{Kim:17-1}, the
division of floating-point numbers (e.g., the division of
${T_{i}(t)-T_{i}(t_{k})}$ by ${1 + \hat{\epsilon}_{i,k}}$ in
\eqref{eq:logical_clock_model}) does not incur much precision loss as the
arithmetic precision supported by most personal computers (PCs) and workstations
is high enough (i.e., 64-bit double-precision floating-point type). For typical
WSN platforms based on a low-cost microcontroller unit (MCU) and limited memory
space, however, floating-point type is generally limited to 32-bit
single-precision, which may result in significant precision loss. As described
in \cite{Dja14:_imp}, implementing high-precision synchronization schemes
requiring floating-point division in WSNs has to be discreet due to the hardware
limitations of the underlying platforms. In case of EE-ASCFR, because the
logical clock updates at sensor nodes in \eqref{eq:logical_clock_model} requires
accurate floating-point division and has a recursive nature, the impact of the
computational errors on the logical clock is accumulated over time.

To avoid the recursive nature of \eqref{eq:logical_clock_model} in EE-ASCFR and
simplify the quantification of the impact of limited precision on time
synchronization, we propose an improved logical clock update equation as
follows:
\begin{equation}
  \label{eq:updated_logical_clock_model}
  \mathscr{T}_{i}\Big(T_{i}(t)\Big) = \mathscr{T}_{i}\Big(T_{i}(t_{0})\Big)
  + \dfrac{T_{i}(t)-T_{i}(t_{0})}{1 + \hat{\epsilon}_{i,k}} ,
\end{equation}
where the current logical clock is updated based on the value of the logical
clock at the first time synchronization, instead of its value at the previous
time synchronization (i.e., $\mathscr{T}_{i}(T_{i}(t_{k}))~$), and the time
duration since the first time synchronization divided by the estimated clock
frequency ratio. Even though there is no recursive term in
\eqref{eq:updated_logical_clock_model}, however, the experiments with the real
testbed show that this improved logical clock update equation still results in
cumulative errors caused by the precision loss involved with the division by
$1 + \hat{\epsilon}_{i,k}$ (i.e., the second term in RHS of
\eqref{eq:updated_logical_clock_model}).

The impact of the precision loss in EE-ASCFR can be analyzed as follows: Because
$\hat{\epsilon}_{i,k}{\ll}1$ in general, the second term in RHS of
\eqref{eq:updated_logical_clock_model} can be approximated by its first-order
Taylor polynomial, i.e.,
\begin{equation}
  \label{eq:taylor_approx}
  \dfrac{T_{i}(t)-T_{i}(t_{0})}{1 + \hat{\epsilon}_{i,k}}  \approx
  \Big(T_{i}(t)-T_{i}(t_{0})\Big) \times (1 - \hat{\epsilon}_{i,k}) .
\end{equation}
Let $\epsilon$ be the precision loss for the clock skew $\hat{\epsilon}_{i,k}$,
i.e.,
\begin{equation}
  \label{eq:precision_loss}
  \epsilon \triangleq \hat{\epsilon}_{i,k} - \hat{\epsilon}_{i,k}^{LP} ,
\end{equation}
where $\hat{\epsilon}_{i,k}^{LP}$ denotes the actual, imprecise value of the
clock skew in implementation due to the limited precision. Then, the
computational error $\Psi$ due to the precision loss can be described as
follows:
\begin{align}
  \label{eq:computation_error}
  \begin{split}
    \Psi & \triangleq \Big(T_{i}(t)-T_{i}(t_{0})\Big) \times (1 -
    \hat{\epsilon}_{i,k}) -
    \Big(T_{i}(t)-T_{i}(t_{0})\Big) \times (1 - \hat{\epsilon}_{i,k}^{LP}) \\
    & = \Big(T_{i}(t)-T_{i}(t_{0})\Big) \times (\hat{\epsilon}_{i,k}^{LP} -
    \hat{\epsilon}_{i,k}) \\
    & = - \Big(T_{i}(t)-T_{i}(t_{0})\Big)\epsilon .
  \end{split}
\end{align}
\eqref{eq:computation_error} shows that, given the precision loss $\epsilon$,
the computational error $\Psi$ is proportional to the time duration since the
first time synchronization. This means that the computational error becomes
larger as the logical clock is continuously updated, because the time duration
since the first time synchronization---i.e.,
$T_{i}(t){-}T_{i}(t_{0})$---increases.

To quantify the impact of the precision loss, we turn back to the definition of
the floating-point formats in IEEE standard 754 \cite{IEEE:754-2008}. According
to the definitions of the 32-bit floating-point and decimal interchange format
parameters in this standard, 7-digit precision is provided for decimal
numbers. As nesC language \cite{nesC}, which is used to build applications on
the TinyOS platform, is basically the extension of the standard C language and
follows the IEEE standard 754 as described in \cite{nesC_float}, the decimal
numbers represented by nesC 32-bit floating-point type provide the limited
precision of 7 digits. Note that the highest clock resolution provided by TinyOS
is \SI{1}{\us}, which limits the precision of all time synchronization schemes
implemented on the TinyOS platform, including EE-ASCFR. Considering the
microsecond-level synchronization limit of the TinyOS and the SI of \SI{10}{\s}
as an example, \SI{e7}{\us} is the actual value involved in the computation of
logical clock in \eqref{eq:logical_clock_model} (i.e., the difference between
the two timestamps). In the worst case, the loss of the precision in the
estimated frequency ratio, whose true value is very close to one, would be
$10^{-7}$ (i.e., $\epsilon$).

\section{Asymmetric High-Precision Time Synchronization (AHTS)}
\label{sec-3}
Based on the results of the investigation of the impact of limited precision
floating-point arithmetic on time synchronization at resource-constrained sensor
nodes with EE-ASCFR in Section~\ref{sec-2}, here we propose AHTS that can
achieve microsecond-level time synchronization accuracy even with
resource-constrained sensor nodes in multi-hop WSNs. We first describe its
system architecture and basic operations and then discuss its extension to
multi-hop topologies.

\subsection{System Architecture and Basic Operations}
\label{sec-3-1}
\begin{figure}[!tb]
  \centering%
  \includegraphics[width=.8\linewidth]{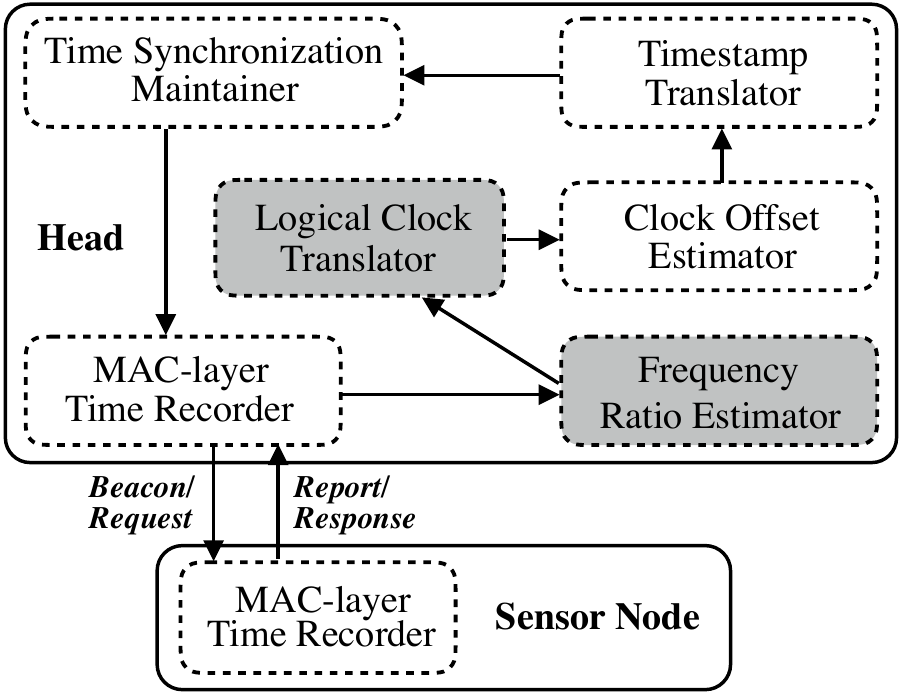}
  \caption{System architecture of AHTS for resource-constrained wireless sensor
    networks.}
  \label{fig:AHTS}
\end{figure}

To address the issues resulting from the limited precision floating-point
arithmetic at resource-constrained sensor nodes, we move all the time
synchronization tasks of sensor nodes except timestamping to the
head\footnote{From now on, we collectively call the head node and the monitoring
  center (i.e., a workstation or a server) connected to it as the head.} in
AHTS: Specifically, as shown in Fig.~\ref{fig:AHTS}, the logical clock
translator described in \eqref{eq:updated_logical_clock_model} and the frequency
ratio estimator (i.e., the cumulative ratio (CR) estimator in
\cite{Kim:13-1})---which run at sensor nodes in EE-ASCFR---are now part of the
time synchronization tasks of the head.

This redistribution of time synchronization tasks between the head and sensor
nodes leaves just timestamping (i.e., the ``MAC-layer Time Recorder'' in
Fig.~\ref{fig:AHTS}) to sensor nodes. As a result, all floating-point arithmetic
operations required by the clock estimation procedures are done at the head with
abundant computing and power resources (including 64-bit double-precision
floating-point arithmetic). Based on this revised system architecture, AHTS
operates as follows:

In the beginning of AHTS operation, the time synchronization maintainer at the
head triggers the time synchronization process, and a hardware clock timestamp
$T1$ is recorded by the MAC-layer time recorder and sent to sensor nodes via a
\textit{Beacon/Request} message. When a sensor node receives the
\textit{Beacon/Request} message, it records the value of its own hardware clock
$T2$. Note that, in the reverse two-way message exchange as implemented in
EE-ASCFR, ${T2_{0}}$ shown in Fig.~\ref{fig:reverse_twoway} is not required by
the head as the estimation of the clock frequency ratio is done at the sensor
node. In AHTS, however, ${T2_{0}}$ is essential for the head to estimate the
clock frequency ratio. Consequently, ${T2_{0}}$ has to be delivered from the
sensor node to the head either through one additional message after the initial
\textit{Beacon/Request} message (i.e., the dotted line in
Fig.~\ref{fig:reverse_twoway}) or embedded in the first \textit{Report/Response}
message later.

When a measurement event occurs, the sensor node records a timestamp $T_{m}$
with respect to its own hardware clock. A \textit{Response/Report} message is
transmitted to the head, carrying the measurement timestamp $T_{m}$ and the most
recently generated $T2$ together with the hardware clock timestamp $T3$ of its
own transmission time. When receiving the \textit{Response/Report} message from
a sensor node, the head records a timestamp $T4$ using its MAC-layer time
recorder. The frequency ratio estimator calculates the clock frequency ratio
based on the differences of current $T2$ and $T1$ to the initial ones (i.e.,
$T2_{0}$ and $T1_{0}$) by employing 64-bit double-precision floating-point
type\footnote{The 64-bit double-precision floating-point type is also named as
  \textit{double} type in common programming languages.}, which has the
precision of 16 digits \cite{IEEE:754-2008}.

Afterwards, the clock offset estimator estimates the clock offset based on the
reverse two-way message exchange as in EE-ASCFR. With the estimated clock
frequency ratio and clock offset, the timestamp translator finally converts the
value of the measurement timestamp $T_{m}$, which is based on the hardware clock
of the sensor node, into that based on the reference clock at the head.

Note that, because AHTS is based on the same reverse two-way message exchange as
EE-ASCFR, it can properly compensate for propagation delay that is ignored in
the time synchronization schemes based on the one-way message dissemination
(e.g., FTSP and RSP). In addition to the propagation delay, the interrupt
delay---i.e., the delay between the transmission and reception interrupts of a
message at a sender and a receiver---is also compensated as part of the two-way
message exchange.

\subsection{Multi-Hop Time Synchronization}
\label{sec-3-2}
In \cite{Kim:17-1}, the extension of EE-ASCFR to a hierarchical structure for
network-wide, multi-hop time synchronization is sketched based on
packet-relaying and time-translating gateways, but no implementation details are
provided. Here we discuss the multi-hop extension of AHTS and the details of its
implementation.

\begin{figure}[!tb]
  \centering%
  \includegraphics[width=\linewidth]{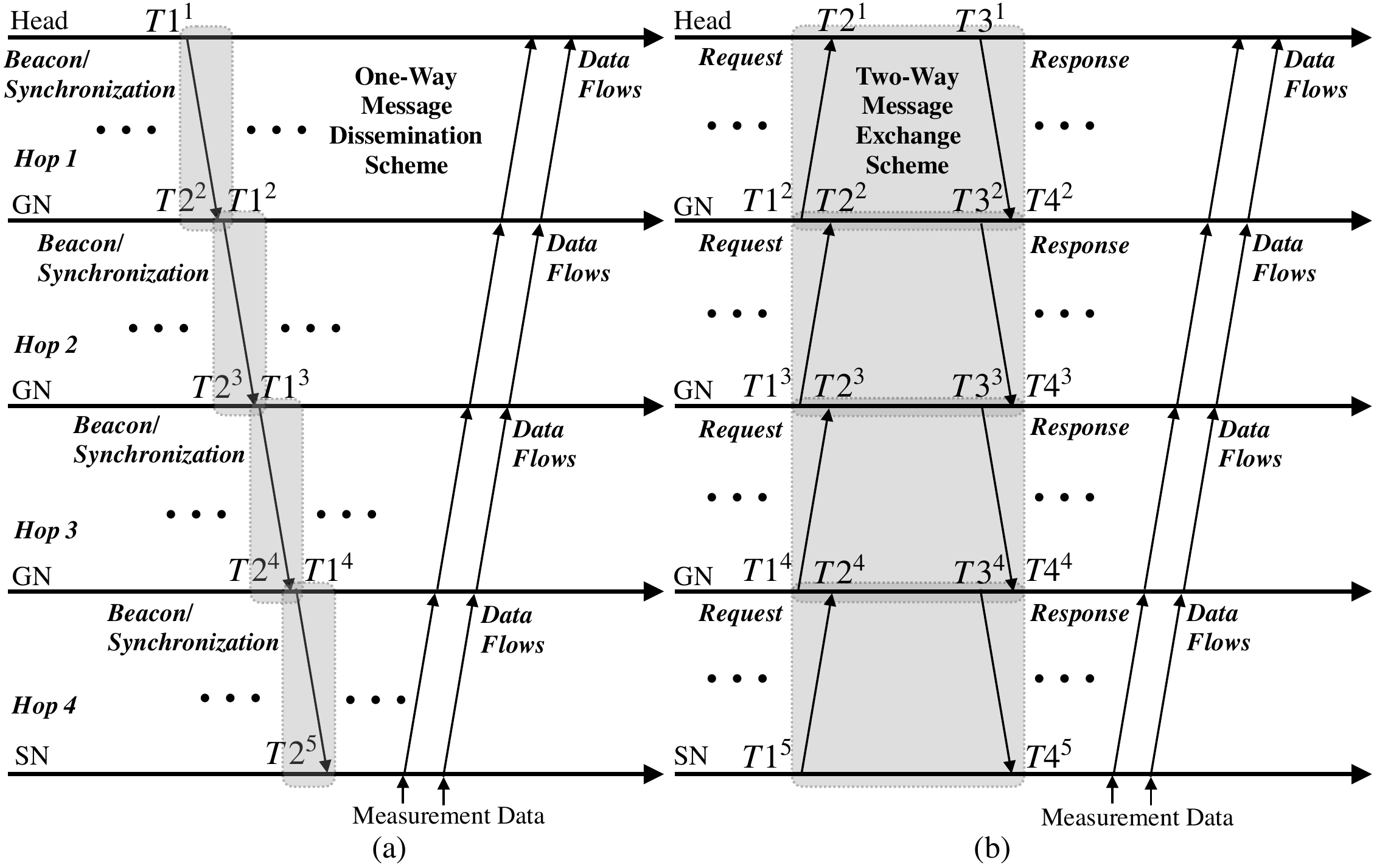}
  \caption{Conventional multi-hop time synchronization schemes based on (a) the
    one-way message dissemination and (b) the two-way message exchange.}
  \label{fig:conventional_multi-hop}
\end{figure}

The conventional multi-hop one-way and two-way time synchronization schemes are
shown in Fig.~\ref{fig:conventional_multi-hop}~(a) and (b), respectively.
As shown in the figure, compared to the time synchronization schemes based on
the one-way message dissemination, those based on the two-way message exchange
(e.g., TPSN) require one additional message at each hop to acquire four
timestamps: For a flat $n$-hop network, for instance, we need $n$
synchronization messages and $2n$ timestamps for the one-way scheme but $2n$
synchronization messages and $4n$ timestamps for the two-way scheme.

In terms of the number of message transmissions, the conventional multi-hop
one-way scheme is more efficient but at the expense of relatively lower time
synchronization accuracy resulting from the lack of propagation delay
compensation. For high-precision time synchronization (e.g., microsecond-level),
by the way, the impact of propagation delay, which is negligible in a single-hop
network, could be accumulated through per-hop forwarding and no longer
negligible in a multi-hop network. Therefore, the propagation delay should be
properly compensated for in the high-precision multi-hop time synchronization
schemes.

\begin{figure}[!tb]
  \centering%
  \includegraphics[width=0.7\linewidth]{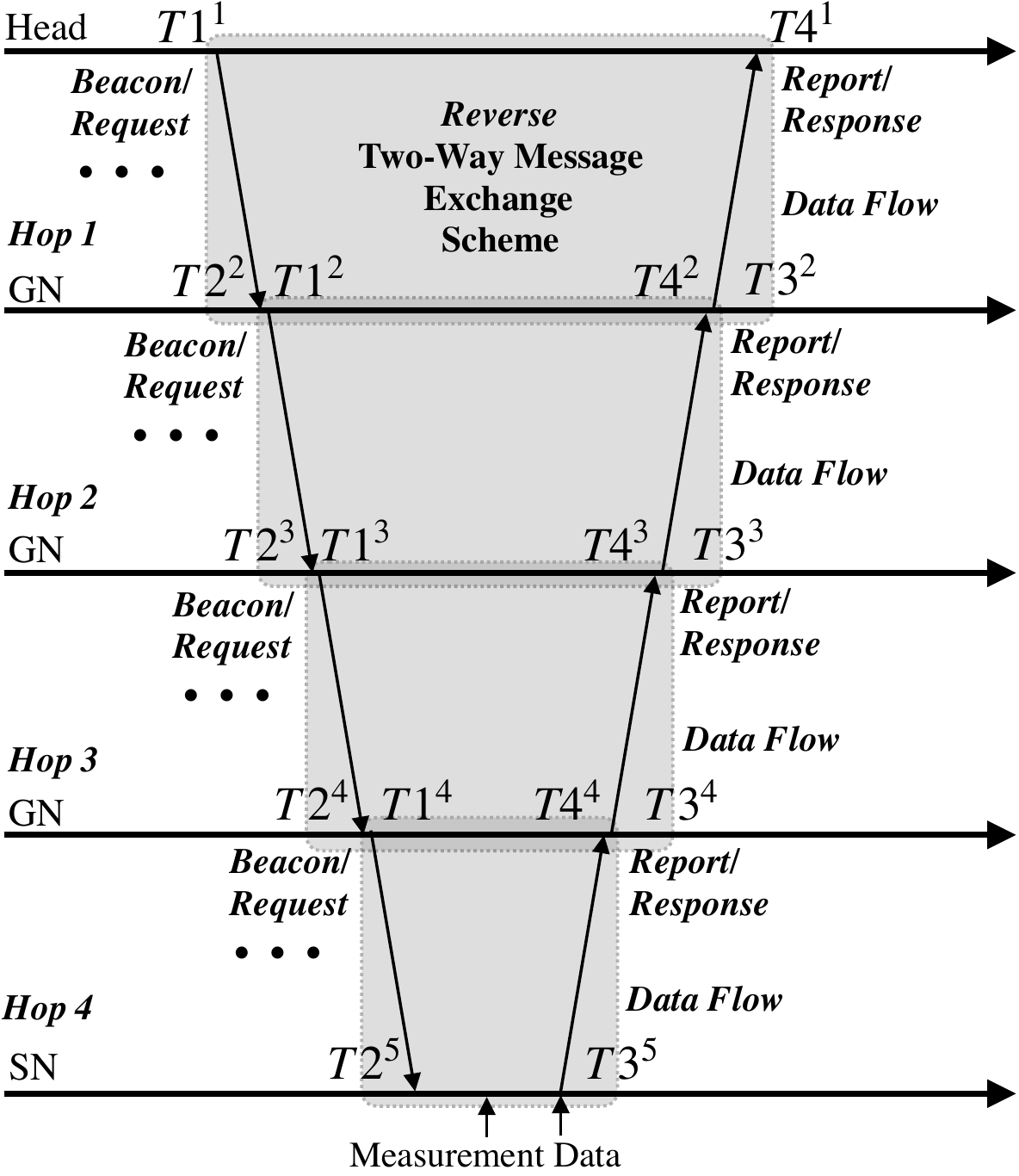}
  \caption{Multi-hop extension of time synchronization schemes based on the
    reverse two-way message exchange with optional bundling of measurement
    data.}
  \label{fig:reverse-two-way_multi-hop}
\end{figure}

Note that both EE-ASCFR and AHTS, i.e., the improved version of EE-ASCFR,
address the issue of the increased number of message transmissions through the
reverse two-way message exchange and the embedment of measurement data in time
synchronization messages, which could reduce the number of message transmissions
to that of the conventional one-way schemes as shown in
Fig.~\ref{fig:reverse-two-way_multi-hop} while compensating for the propagation
delay as in the conventional two-way schemes. In addition, both EE-ASCFR and
AHTS could further reduce the number of message transmissions through
measurement data bundling, which is also shown in
Fig.~\ref{fig:reverse-two-way_multi-hop}.

As for the processing of the timestamps related with the reverse two-way message
exchange over multiple hops, we adopt the time-translation approach described in
\cite{Kim:17-1} but again move its processing from the intermediate gateway
nodes to the head in order to address the increased energy consumption and
processing at the gateway nodes that are likely to be battery-powered like other
sensor nodes. Specifically, each hop of the multi-hop network maintains the same
timestamping procedure as in the single-hop network (e.g., $T1^{3}$, $T2^{4}$,
$T3^{4}$, and $T4^{3}$ for Hop~3, and $T1^{4}$, $T2^{5}$, $T3^{5}$, and $T4^{4}$
for Hop~4 in Fig.~\ref{fig:reverse-two-way_multi-hop}). Then the upper
node---i.e., the node working as a gateway for its lower nodes in the original
time-translating gateway approach in EE-ASCFR---just transfers the set of the
four collected timestamps to the head, which eventually handles the translation
of the measurement times embedded in a packet with the timestamps based on the
logical clocks and the offsets for the two nodes.

In this way, we eliminate the impact of any extra delays on time synchronization
such as packet delays resulting from queueing and MAC operations, which are
accumulated through per-hop forwarding in multi-hop networks and could severely
affect the other approach based on packet relaying gateway nodes as discussed in
\cite{Kim:17-1}.

\section{Experimental Results}
\label{sec-4}
We carry out a comparative analysis of the time synchronization performance of
EE-ASCFR and the proposed AHTS based on a series of experiments for both
single-hop and multi-hop scenarios with a real WSN testbed consisting of TelosB
motes running TinyOS.

As discussed in Section~\ref{sec-2}, the implementation of EE-ASCFR on a real
WSN testbed reveals the significant impact of the limited precision
floating-point arithmetic of resource-constrained sensor nodes on time
synchronization performance. The focus of the experiments and their analyses,
therefore, is put on how the proposed AHTS addresses the issue of the precision
loss resulting from the use of single-precision floating-point format at
resource-constrained sensor nodes in time synchronization.

\subsection{Single-Hop Scenario}
\label{sec-4-1}
\begin{figure}[!tb]
  \newlength{\myvspacea} \setlength{\myvspacea}{0.1cm}
  \begin{center}
    \includegraphics[width=.7\linewidth]{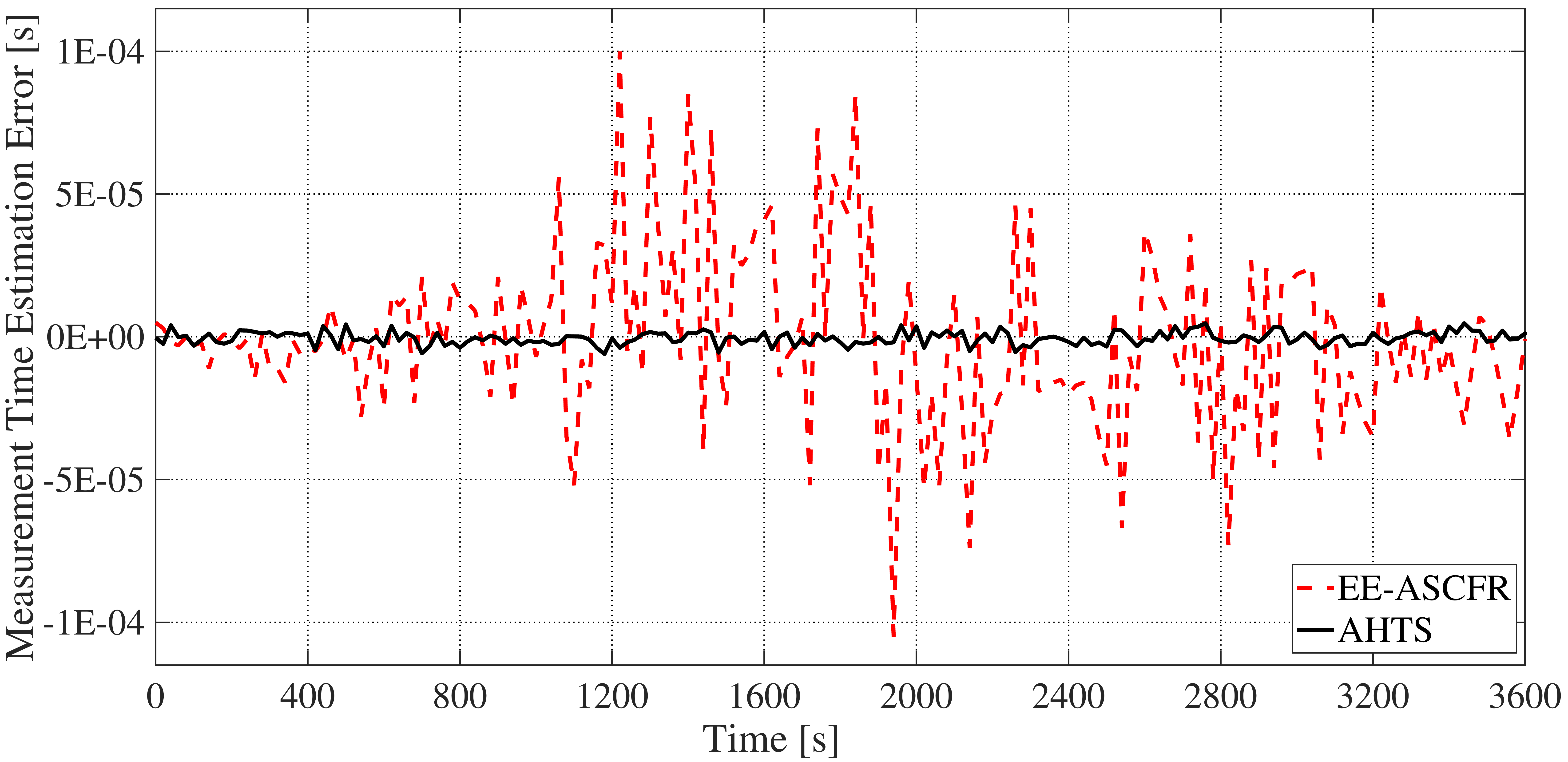} \\
    {\scriptsize (a)}
    \vspace{\myvspacea}\\
    \includegraphics[width=.7\linewidth]{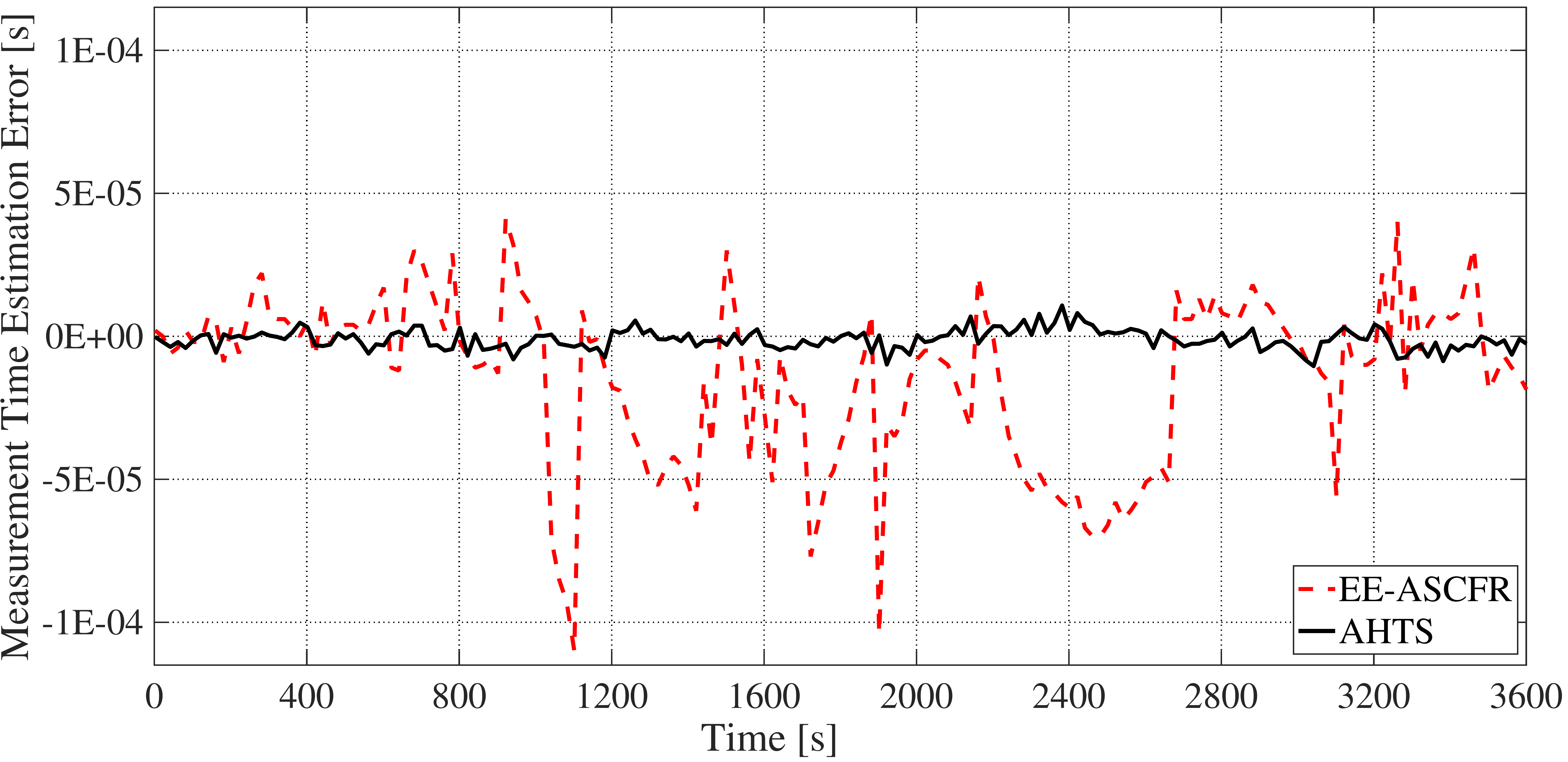} \\
    {\scriptsize (b)}
    \vspace{\myvspacea}\\
    \includegraphics[width=.7\linewidth]{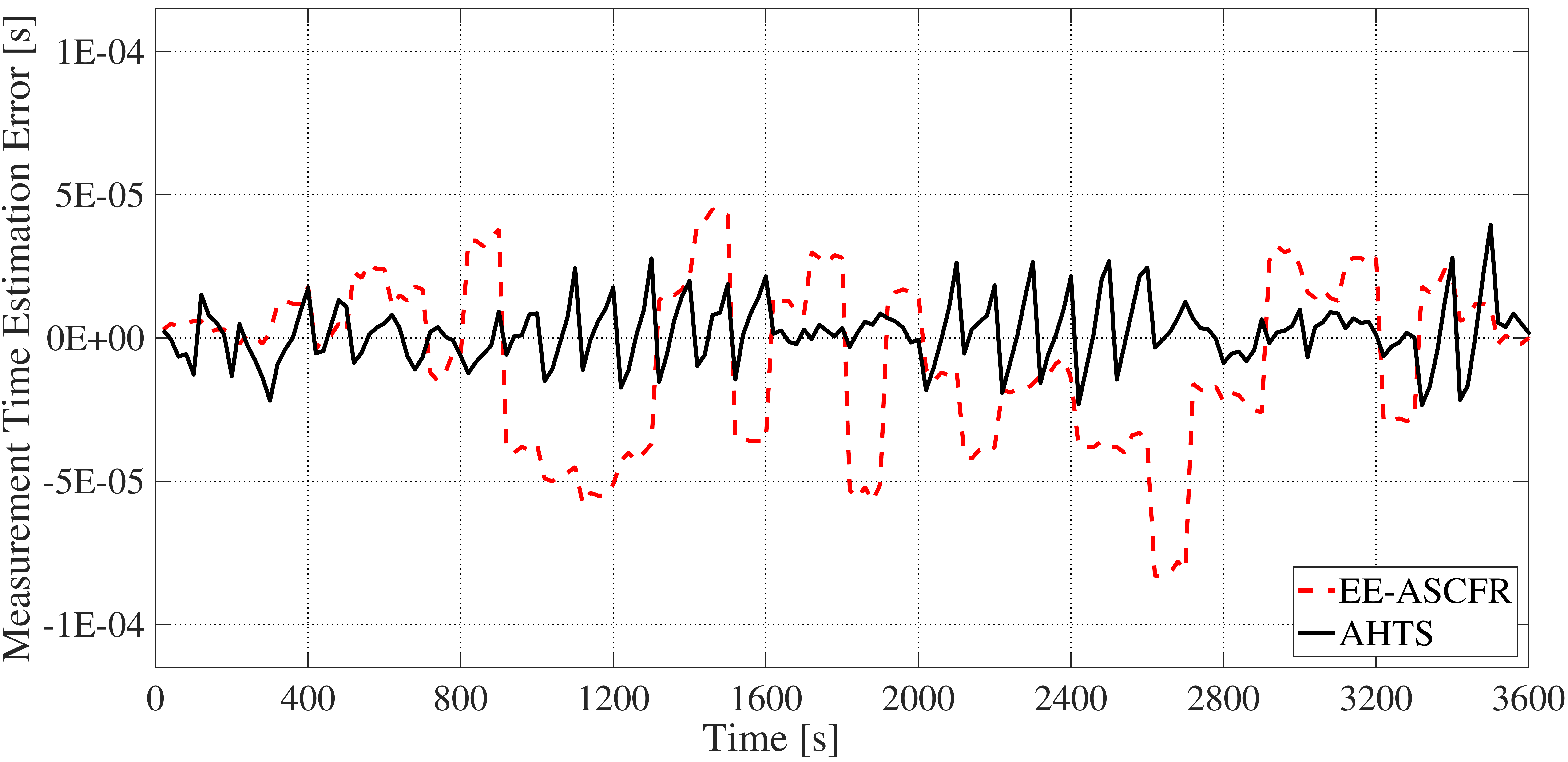}\\
    {\scriptsize (c)}
  \end{center}
  \caption{Measurement time estimation errors of EE-ASCFR and AHTS with SI of
    (a) \SI{1}{\s}, (b) \SI{10}{\s} and (c) \SI{100}{\s} for the single-hop
    scenario.}
  \label{fig:MTESoverTime}
\end{figure}

First, we consider a single-hop WSN with one head and one sensor node. The
experiments are run over a period of \SI{3600}{\s} with three different values
of SI, i.e., \SI{1}{\s}, \SI{10}{\s} and \SI{100}{\s}. The number of bundled
measurements is set to 5 for all the experiments. Fig.~\ref{fig:MTESoverTime}
shows the measurement time estimation errors of EE-ASCFR and AHTS, and
Table~\ref{tab:time_sync_results_si} summarizes the mean absolute error (MAE)
and the mean squared error (MSE) of their measurement time estimation.
\begin{table}[!tb]
  \centering
  \begin{threeparttable}
    \caption{MAE and MSE of Measurement Time Estimation of EE-ASCFR and AHTS for
      the Single-Hop Scenario}
    \label{tab:time_sync_results_si}
    \centering \setlength{\tabcolsep}{6mm}
   	\begin{tabular}{|c|l||r|r|}
      \hline
      \multicolumn{2}{|c||}{Synchronization Scheme}
      & \multicolumn{1}{c|}{MAE \tnote{1}}
      & \multicolumn{1}{c|}{MSE \tnote{1}} \\ \hline\hline
      \multirow{3}{*}{EE-ASCFR}
      & SI$\;=100~\mbox{s}$  & 2.7276E-05 & 1.0391E-09 \\ \cline{2-4}
      & SI$\;=10~\mbox{s}$  & 2.5182E-05 & 1.1559E-09  \\ \cline{2-4}
      & SI$\;=1~\mbox{s}$  & 2.4069E-05 & 1.0095E-09 \\ \hline\hline
      \multirow{3}{*}{AHTS}
      & SI$\;=100~\mbox{s}$  & 8.4225E-06 & 1.2524E-10 \\ \cline{2-4}
      & SI$\;=10~\mbox{s}$  & 2.3385E-06 & 9.1694E-12  \\ \cline{2-4}
      & SI$\;=1~\mbox{s}$  & 1.8166E-06 & 5.2094E-12  \\ \hline
    \end{tabular}
    \begin{tablenotes}
    \item[1] The samples measured between \SI{360}{\s} (i.e., a tenth of the
      total observation period) and \SI{3600}{\s} are employed to avoid the
      effect of a transient period.
    \end{tablenotes}
  \end{threeparttable}
\end{table}

The results of Fig.~\ref{fig:MTESoverTime} and
Table~\ref{tab:time_sync_results_si} show that the measurement time
synchronization errors of AHTS are stable and much smaller than those of
EE-ASCFR over the observation period for all three values of SI, which
demonstrates that the proposed AHTS successfully addresses the issue of
precision loss in the logical clock update discussed in Section~\ref{sec-2}.

The effect of SI on time synchronization is more visible in AHTS than EE-ASCFR,
especially for the SI of \SI{100}{\s}, which may result from a larger range of
clock drift over a longer period of time by the sensor node's hardware clock
based on a cheap quartz crystal oscillator. In case of EE-ASCFR, the effect of
SI becomes less visible as time goes on (i.e., over \SI{800}{\s}), because it is
overshadowed by that of the aforementioned precision loss.

Overall, the experimental results from the single-hop scenario show that the
proposed AHTS successfully addresses the issue of the precision loss in time
synchronization at resource-constrained sensor nodes and can deliver
microsecond-level time synchronization accuracy.

\subsection{Multi-Hop Scenario}
\label{sec-4-2}
\begin{figure}[!tb]
  \centering
  \includegraphics[width=.9\linewidth]{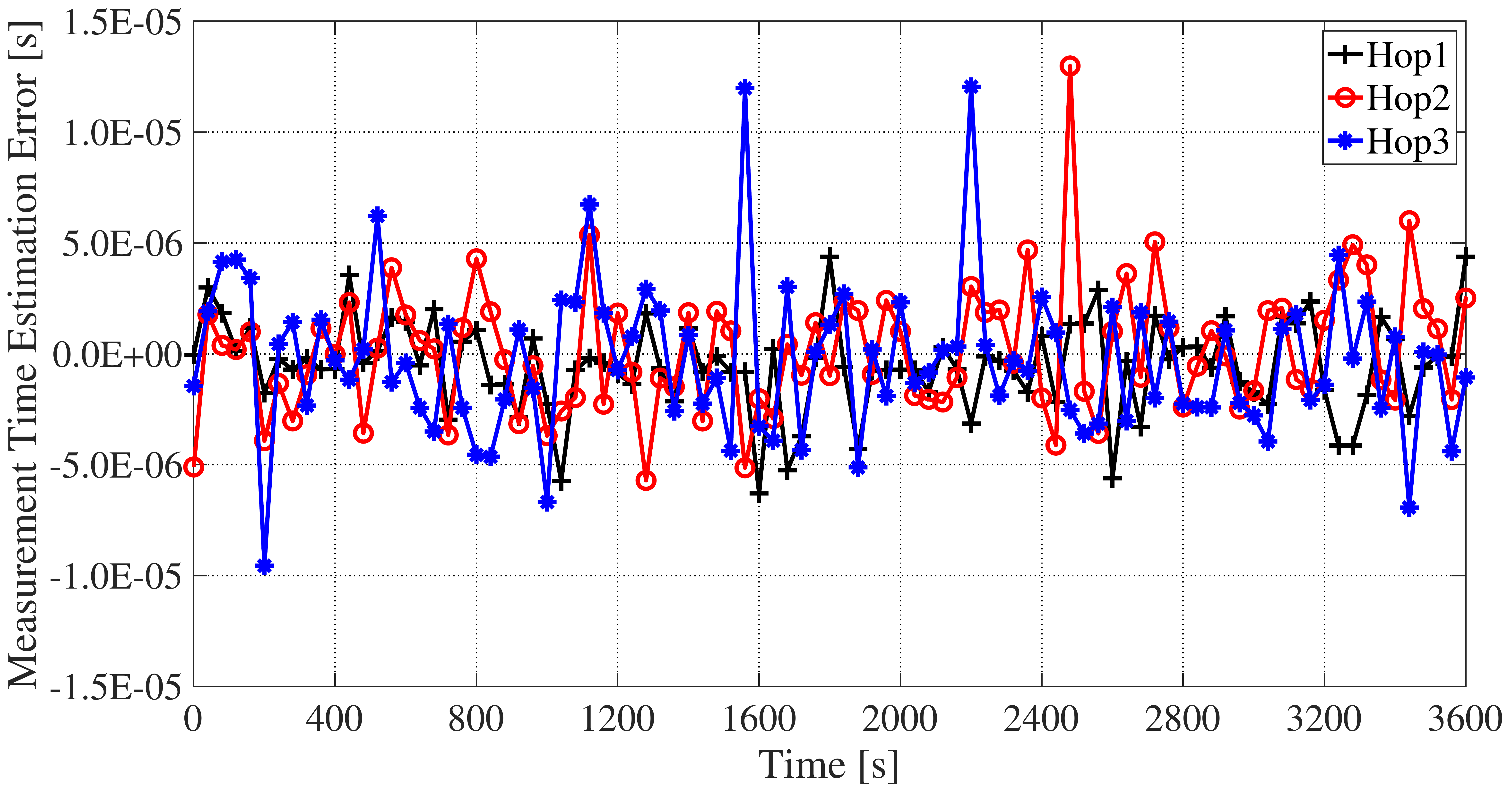} \\
  \caption{Measurement time estimation errors of AHTS with SI of \SI{1}{\s} for
    the multi-hop scenario.}
  \label{fig:evaluation_multihop_line}
\end{figure}

To investigate the effect of the number of hops on time synchronization in
multi-hop topologies, we also consider a multi-hop WSN with one head and three
sensor nodes. The experiments are run over a period of \SI{3600}{\s} as in the
single-hop scenario, but the SI value is fixed to \SI{1}{\s}. For a fair
comparison of the time synchronization performance of sensor nodes in the
multi-hop scenario, each sensor node bundles only its own measurement and
synchronization data with the number of bundled measurements set to 5 as in the
single-hop scenario. Fig.~\ref{fig:evaluation_multihop_line} shows the
measurement time estimation errors of AHTS for different number of hops, and
Table~\ref{tab:time_sync_results_multihop} summarizes the MAE and the MSE of the
measurement time estimation.
\begin{figure}[!tb]
  \centering
  \includegraphics[width=.9\linewidth]{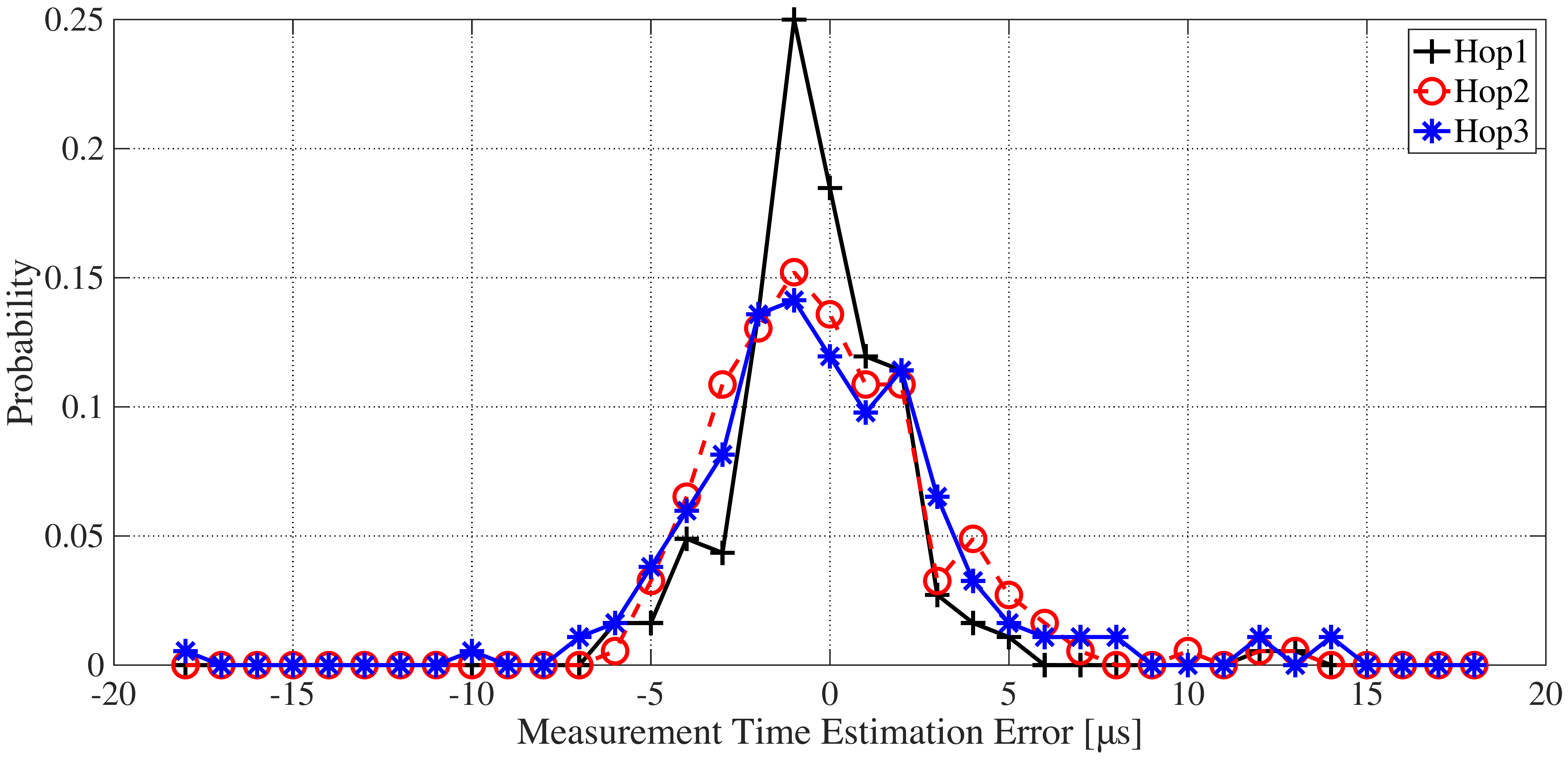} \\
  \caption{Probability distribution of the measurement time estimation errors of
    AHTS with SI of \SI{1}{\s} for the multi-hop scenario.}
  \label{fig:evaluation_multihop_dis}
\end{figure}
\begin{table}[!tb]
  \centering
  \begin{threeparttable}
    \caption{MAE and MSE of Measurement Time Estimation of AHTS for the
      Multi-Hop Scenario}
    \label{tab:time_sync_results_multihop}
    \centering
    
    \setlength{\tabcolsep}{8mm}
    
    \begin{tabular}{|c|l||r|r|}
      \hline
      \multicolumn{2}{|c||}{Hop Number}
      & \multicolumn{1}{c|}{MAE \tnote{1}}
      & \multicolumn{1}{c|}{MSE \tnote{1}} \\ \hline\hline
      \multirow{3}{*}{Hop}
      & $3$  & 2.5700E-06 & 1.2432E-11 \\ \cline{2-4}
      & $2$  & 2.3648E-06 & 1.1056E-11 \\ \cline{2-4}
      & $1$  & 2.1774E-06 & 9.4104E-12 \\ \hline
    \end{tabular}

    \begin{tablenotes}
    \item[1] The samples measured between \SI{360}{\s} (i.e., a tenth of the
      observation period) and \SI{3600}{\s} are employed to avoid the effect of
      a transient period.
    \end{tablenotes}
  \end{threeparttable}
\end{table}

From Table~\ref{tab:time_sync_results_multihop}, we found that the MAE of
measurement time estimation for Hop~1 is \SI{2.1774}{\us}, which is close to
that of the single-hop scenario (i.e., \SI{1.8166}{\us}). We also found that,
due to the layer-by-layer translation of the multi-hop extension, the MAE of
measurement time estimation slightly increases as the hop count increases, which
amounts to about \SI{0.2}{\us} per hop. Likewise, the measurement time
estimation errors in Fig.~\ref{fig:evaluation_multihop_line} show more
fluctuations for Hop~2 and Hop~3 than Hop~1.

Similar behaviors are also observed in the distributions of the measurement time
estimation errors shown in Fig.~\ref{fig:evaluation_multihop_dis}.
From the figure, we can find that about 18\% of the measurement time estimation
errors for Hop~1 are close to zero, while this percentage of time estimation
errors close to zero decreases to around 14\% and 13\% for Hop~2 and Hop~3,
respectively. Note that most of the measurement time estimation errors are
within the range of \SI{-10}{\us} and \SI{10}{\us} for all hop counts in
Fig.~\ref{fig:evaluation_multihop_dis}, which demonstrates that the proposed
AHTS could provide high-precision time synchronization in multi-hop networks as
well as single-hop networks.

\section{Conclusions}
\label{sec-5}
In this paper, we have investigated the actual performance of EE-ASCFR proposed
in \cite{Kim:17-1}---i.e., the state-of-the-art propagation delay and clock skew
compensated time synchronization scheme designed to provide
sub-microsecond-level synchronization accuracy---on resource-constrained sensor
nodes and, based on the results of the investigation with a real WSN testbed,
proposed AHTS to address the issues raised from the practical implementation of
EE-ASCFR, which can achieve high-precision network-wide time synchronization
even with resource-constrained sensor nodes in multi-hop topologies. Noting that
the limited precision in floating-point arithmetic of the resource-constrained
sensor nodes (i.e., 32-bit single-precision on a broad range of WSN platforms
including motes running TinyOS and Arduino) has negative impacts on the time
synchronization performance of EE-ASCFR, we move all the time synchronization
tasks of sensor nodes except timestamping to the head in AHTS and apply this
approach to its extension to multi-hop topologies as well.

The results of experiments for a single-hop scenario with a real WSN testbed
consisting of TelosB motes running TinyOS show that the proposed AHTS
consistently outperforms EE-ASCFR in terms of measurement time estimation errors
by successfully addressing the issue of the precision loss in time
synchronization at resource-constrained sensor nodes and can deliver
microsecond-level time synchronization accuracy with all three values of SI
considered---i.e., \SI{1}{\s}, \SI{10}{\s} and \SI{100}{\s}. We have also
carried out experiments for a three-hop WSN consisting of one head and three
sensor nodes to investigate the effect of the number of hops on the time
synchronization performance of AHTS in multi-hop topologies. The results show
that the MAE of measurement time estimation for Hop~1 is \SI{2.1774}{\us}, which
is close to that of the single-hop scenario (i.e., \SI{1.8166}{\us}), and that
the MAE of measurement time estimation increases as the hop count increases by
around \SI{0.2}{\us} per hop. The distributions of the measurement time
estimation errors from the multi-hop experiments show that most of the
measurement time estimation errors are within the range of \SI{-10}{\us} and
\SI{10}{\us} for all hop counts, which demonstrates that the proposed AHTS could
provide high-precision time synchronization in multi-hop networks as well as
single-hop networks.

Related with our investigation in this paper on the actual performance of
high-precision propagation delay and clock skew compensated time synchronization
schemes on resource-constrained sensor nodes and addressing the issues raised
from their practical implementation on a real WSN testbed, we have identified
several areas of further investigation.

First, the scalability of AHTS needs to be studied on a testbed with a larger
number of sensor nodes. Because there are lots of messages for not only
timestamps but also measurements to be exchanged between the head and sensor
nodes in multi-hop topologies, the bundling of measurement messages in a limited
size of the payload of synchronization messages at both end and intermediate
sensor nodes and their impact on the overall network traffic are to be carefully
studied.

Second, as the proposed AHTS is based on EE-ASCFR designed for high
energy-efficiency as well as high-precision time synchronization, the energy
consumption of AHTS is to be examined with the actual measurement on a real
testbed.

Note that the underlying design assumption of having a more resourceful head
node in AHTS well suits to not only a variety of WSN applications but also
future IoT deployments \cite{badihi18:_time_iot}.

%

\section*{Acknowledgment}
This work was supported by Xi'an Jiaotong-Liverpool University Research
Development Fund (RDF) under grant reference number RDF-16-02-39.

\bibliographystyle{IEEEtran}%
\bibliography{IEEEabrv,kks}%

\end{document}